\documentclass[sigplan,screen]{acmart}
\AtBeginDocument{%
  \providecommand\BibTeX{{%
    \normalfont B\kern-0.5em{\scshape i\kern-0.25em b}\kern-0.8em\TeX}}}

\setcopyright{acmcopyright}
\copyrightyear{2023}
\acmYear{2023}
\setcopyright{acmcopyright}\acmConference[HPC ASIA 2023]{International Conference on High Performance Computing in Asia-Pacific Region}{February 27-March 2, 2023}{Singapore, Singapore}
\acmBooktitle{International Conference on High Performance Computing in Asia-Pacific Region (HPC ASIA 2023), February 27-March 2, 2023, Singapore, Singapore}
\acmPrice{15.00}
\acmDOI{10.1145/3578178.3578238}
\acmISBN{978-1-4503-9805-3/23/02}

\usepackage{booktabs}
\usepackage{blkarray}
\usepackage{graphicx}
\usepackage{multirow}
\usepackage{url}
\usepackage{mathtools}
\usepackage{amsmath}
\usepackage{footnote}
\usepackage{enumitem}
\usepackage{caption}
\usepackage{listings}
\usepackage{xcolor}
\usepackage{parcolumns}

\definecolor{mGreen}{rgb}{0,0.6,0}
\definecolor{mGray}{rgb}{0.5,0.5,0.5}
\definecolor{mPurple}{rgb}{0.58,0,0.82}
\definecolor{red}{rgb}{1, 0, 0}
\definecolor{backgroundColour}{rgb}{0.95,0.95,0.92}

\definecolor{verylightgray}{gray}{0.9}
\lstdefinestyle{CStyle}{
    backgroundcolor=\color{gray!8},
    commentstyle=\color{mGreen!90},
    keywordstyle=\color{blue!100},
    numberstyle=\tiny\color{purple!100},
    stringstyle=\color{black},
    basicstyle=\footnotesize,
    breakatwhitespace=false,         
    breaklines=true,                 
    captionpos=b,                    
    keepspaces=true,                 
    numbers=left,                    
    numbersep=5pt,                  
    showspaces=false,                
    showstringspaces=false,
    showtabs=false,                  
    tabsize=2,
    language=C
}

\begin{document}

%%
%% The "title" command has an optional parameter,
%% allowing the author to define a "short title" to be used in page headers.
\title{Reducing shared memory footprint to leverage high throughput on Tensor Cores and its flexible API extension library}

%%
%% The "author" command and its associated commands are used to define
%% the authors and their affiliations.
%% Of note is the shared affiliation of the first two authors, and the
%% "authornote" and "authornotemark" commands
%% used to denote shared contribution to the research.
\author{Hiroyuki Ootomo}
\affiliation{%
  \institution{Tokyo Institute of Technology}
  \city{Tokyo}
  \country{Japan}}
\email{ootomo.h@rio.gsic.titech.ac.jp}
\author{Rio Yokota}
\affiliation{%
  \institution{Tokyo Institute of Technology}
  \city{Tokyo}
  \country{Japan}}

%\author{Aparna Patel}
%\affiliation{%
% \institution{Rajiv Gandhi University}
% \streetaddress{Rono-Hills}
% \city{Doimukh}
% \state{Arunachal Pradesh}
% \country{India}}
%
%\author{Huifen Chan}
%\affiliation{%
%  \institution{Tsinghua University}
%  \streetaddress{30 Shuangqing Rd}
%  \city{Haidian Qu}
%  \state{Beijing Shi}
%  \country{China}}
%
%\author{Charles Palmer}
%\affiliation{%
%  \institution{Palmer Research Laboratories}
%  \streetaddress{8600 Datapoint Drive}
%  \city{San Antonio}
%  \state{Texas}
%  \country{USA}
%  \postcode{78229}}
%\email{cpalmer@prl.com}
%
%\author{John Smith}
%\affiliation{%
%  \institution{The Th{\o}rv{\"a}ld Group}
%  \streetaddress{1 Th{\o}rv{\"a}ld Circle}
%  \city{Hekla}
%  \country{Iceland}}
%\email{jsmith@affiliation.org}
%
%\author{Julius P. Kumquat}
%\affiliation{%
%  \institution{The Kumquat Consortium}
%  \city{New York}
%  \country{USA}}
%\email{jpkumquat@consortium.net}

%%
%% By default, the full list of authors will be used in the page
%% headers. Often, this list is too long, and will overlap
%% other information printed in the page headers. This command allows
%% the author to define a more concise list
%% of authors' names for this purpose.
\renewcommand{\shortauthors}{Ootomo and Yokota}

%%
%% The abstract is a short summary of the work to be presented in the
%% article.
\begin{abstract}
Matrix-matrix multiplication is used for various linear algebra algorithms such as matrix decomposition and tensor contraction.
NVIDIA Tensor Core is a mixed-precision matrix-matrix multiplication and addition computing unit, where the theoretical peak performance is more than 300 TFlop/s on NVIDIA A100 GPU.
NVIDIA provides WMMA API for using Tensor Cores in custom kernel functions.
The most common way to use Tensor Core is to supply the input matrices from shared memory, which has higher bandwidth than global memory.
However, the Bytes-per-Flops (B/F) ratio of the shared memory and Tensor Cores is small since the performance of Tensor Cores is high.
Thus, it is important to reduce the shared memory footprint for efficient Tensor Cores usage.
In this paper, we analyze the simple matrix-matrix multiplication on Tensor Cores by the roofline model and figure out that the bandwidth of shared memory might be a limitation of the performance when using WMMA API.
To alleviate this issue, we provide a WMMA API extension library to boost the throughput of the computation, which has two components.
The first one allows for manipulating the array of registers input to Tensor Cores flexibly.
We evaluate the performance improvement of this library.
The outcome of our evaluation shows that our library reduces the shared memory footprint and speeds up the computation using Tensor Cores.
The second one is an API for the SGEMM emulation on Tensor Cores without additional shared memory usage.
We have demonstrated that the single-precision emulating batch SGEMM implementation on Tensor Cores using this library achieves 54.2 TFlop/s on A100 GPU, which outperforms the theoretical peak performance of FP32 SIMT Cores while achieving the same level of accuracy as cuBLAS.
The achieved throughput can not be achieved without reducing the shared memory footprint done by our library with the same amount of register usage.
\end{abstract}

%%
%% The code below is generated by the tool at http://dl.acm.org/ccs.cfm.
%% Please copy and paste the code instead of the example below.
%%
\begin{CCSXML}
<ccs2012>
   <concept>
       <concept_id>10011007.10011006.10011072</concept_id>
       <concept_desc>Software and its engineering~Software libraries and repositories</concept_desc>
       <concept_significance>500</concept_significance>
       </concept>
 </ccs2012>
\end{CCSXML}

\ccsdesc[500]{Software and its engineering~Software libraries and repositories}

%%
%% Keywords. The author(s) should pick words that accurately describe
%% the work being presented. Separate the keywords with commas.
\keywords{Tensor Cores, WMMA API, GPU}

%% A "teaser" image appears between the author and affiliation
%% information and the body of the document, and typically spans the
%% page.
\received{20 February 2007}
\received[revised]{12 March 2009}
\received[accepted]{5 June 2009}

%%
%% This command processes the author and affiliation and title
%% information and builds the first part of the formatted document.
\maketitle

\section{Introduction}

NVIDIA Tensor Core is a mixed-precision matrix multiplication and addition computing unit with up to 312 TFlop/s on NVIDIA A100 GPU \cite{corporation_nvidia_2022}.
From the demand for high-throughput matrix multiplication from deep learning, several computing units specialized for matrix multiplication are developed, such as Google TPU \cite{jouppi_-datacenter_2017}, AMD Matrix Core, Intel Ponte Vecchio, and Preferred Networks MN-Core \cite{makino_near-optimal_2021}.
Tensor Core computes the multiplication of two matrices where the data type is low-precision in high throughput and high-precision.
Although Tensor Core is developed for deep learning, especially fully-connected layer and convolution layer computations, it is applied to other fields of computations and fundamental linear algebra algorithms leveraging the low- and mixed-precision feature \cite{haidar_harnessing_2018,dakkak_accelerating_2019,li_tcfft_2021,finkelstein_quantum_2022,markidis_nvidia_2018,ootomo_recovering_2022}.
NVIDIA provides highly optimized libraries for using Tensor Cores which can be called from a host, such as cuBLAS and cuDNN.
We can leverage the high throughput of Tensor Core using these libraries without special knowledge of it.
Furthermore, NVIDIA also provides an API for use inside a CUDA kernel function called WMMA (Warp Matrix Multiply Accumulate) API.
This API provides basic functionalities such as loading matrix data from memory, multiplication and addition on Tensor Core, and storing the resulting matrix data in memory.
Using this API, we load matrix data from the device memory or shared memory to an array of registers called ``fragment'' to input Tensor Cores.
On the other hand, there are some matrices where each element can be computed on the fly, for instance, the Householder matrix and Given's rotation matrix.
Even for these matrices, we have to store them in memory and load them since the API is too simple and lacks flexibility, and this can degrade the throughput.
Thus, for instance, Dakkak \textit{et al.} \cite{dakkak_accelerating_2019} use Tensor Cores for reduction and scan operation by generating a fragment of an upper triangular matrix and a lower triangular matrix without generating the matrices on the shared memory.
Li \textit{et al.} \cite{li_tcfft_2021} use Tensor Cores in FFT operations by generating fragments directly.
However, since NVIDIA does not provide information on fragment mapping, we need to analyze the structure of the fragment by ourselves to generate the fragment in these ways.
For another example, the single-precision matrix-matrix multiplication emulation method on Tensor Cores \cite{ootomo_recovering_2022} accesses the shared memory more than necessary if we only use the WMMA API.
Therefore, the throughput of the emulation method can degrade if we only use the API.

In this paper, we first show that it is important to reduce the shared memory footprint to leverage the high Tensor Cores performance.
We analyze a matrix-matrix multiplication on Tensor Cores using the roofline model \cite{williams_roofline_2009}.
As a result, it is difficult to leverage the high Tensor Cores performance without sufficient register blocking or reducing the shared memory footprint.
However, the number of registers is limited.
To reduce the shared memory footprint, we implement a WMMA API extension library, which flexibly manipulates the input register array of Tensor Cores by analyzing the memory and register array mappings.
This library can generate an arbitrary input register array without an extra shared memory footprint.
Furthermore, we provide an API for single-precision matrix-matrix multiplication emulation on Tensor Cores, which has the same interface as WMMA API.
Our goals of this work are 1) to reveal that the shared memory bandwidth can degrade the utilization efficiency of Tensor Cores in some cases and 2) to provide a library to reduce such degradation by manipulating the fragment flexibly for reducing the shared memory footprint.

Our contributions are as follows:
\begin{itemize}
    \item
    We show that the shared memory bandwidth might limit the matrix-matrix multiplication performance on Tensor Cores by roofline model analysis.
    Furthermore, we find it important to reduce the shared memory footprint on NVIDIA A100 compared to V100 since the Bytes-per-Flops (B/F) ratio of the Tensor Core performance and shared memory bandwidth on NVIDIA A100 is smaller than V100.
    %%%%%Furthermore, we find that the performance on NVIDIA A100 Tensor Cores are more affected by the shared memory bandwidth compared to V100.%%%%%%
    \item
    We implement a general WMMA API extension library to reduce the shared memory footprint.
    By using this library, we can manipulate the fragment elements flexibly.
    And as a secondary effect, we can reduce the shared memory usage in some cases since some of the temporary shared memory areas for generating matrices that are loaded as fragments become unnecessary.
    We investigate the availability of this library and find the condition to speed up the fragment generation.
    The library is available on GitHub\footnote{\url{https://github.com/wmmae/wmma_extension}}.
    \item
    We figure out that by the inflexibility of WMMA API, shared memory bandwidth bounds the theoretical peak performance of single-precision matrix-matrix multiplication emulation on Tensor Cores.
    By using our extension library, we improve its theoretical peak performance.
    Furthermore, we provide functionality for that which has the same interface as WMMA API.
    To demonstrate the usability of the functionality, we implement batched matrix-matrix multiplication using the functionality.
    We show that our implementation outperforms the FP32 theoretical peak performance on NVIDIA A100 while the accuracy is the same level as cuBLAS SGEMM.
\end{itemize}
%Our library is available on GitHub: \url{xxxx}.

\section{Background}

\subsection{Shared memory}
\subsubsection{The bandwidth of shared memory}

\begin{table}[]
\center
\begin{tabular}{r|cccc}
\toprule
                      & V100                 & V100S                & \multicolumn{2}{c}{A100}                    \\
                      & (SXM2)               & (PCIe)               & \multicolumn{2}{c}{(SXM4/PCIe)}                  \\
\midrule
\midrule
SMs                   & \multicolumn{2}{c}{80}                      & \multicolumn{2}{c}{108}                     \\
Clock {[}MHz{]}       & 1,380                & 1,597                & \multicolumn{2}{c}{1,410}                \\
\midrule
{\bf Device memory}   &                      &                      &                      &                      \\
Size {[}GB{]}         & 32/16                & 32                   & 40                   & 80                   \\
Bandwidth {[}GB/s{]}  & 900                  & 1,134                & 1,555                & 2,039                \\
\midrule
{\bf Shared memory}   &                      &                      &                      &                      \\
Size {[}KB/SM{]}      & \multicolumn{2}{c}{$\sim$96}                & \multicolumn{2}{c}{$\sim$164}               \\
Bandwidth {[}GB/s{]}  & 14,131               & 16,353               & \multicolumn{2}{c}{19,491}                  \\
\midrule
{\bf Performance}     & \multicolumn{1}{l}{} & \multicolumn{1}{l}{} & \multicolumn{1}{l}{} & \multicolumn{1}{l}{} \\
FP32 {[}TFlop/s{]}    & 15.7                 & 16.4                 & \multicolumn{2}{c}{19.5}                    \\
FP16 {[}TFlop/s{]}    & 31.4                 & 32.8                 & \multicolumn{2}{c}{39.0}                    \\
FP16-TC {[}TFlop/s{]} & 112                  & 125                  & \multicolumn{2}{c}{312}                     \\
TF32-TC {[}TFlop/s{]} & -                    & -                    & \multicolumn{2}{c}{156}                     \\
\bottomrule
\end{tabular}
\caption{Specifications of NVIDIA GPUs A100 and V100.}
\label{tab:gpu-spec}
\end{table}
The shared memory is a high bandwidth, low latency, and small size compared to the device memory.
This memory is located on each Streaming Multiprocessor (SM) and shared by all threads in a thread block.
The shared memory is divided into the same size memory modules called banks.
In CUDA, a cluster of threads consisting of 32 threads is called a warp, and when multiple threads in a warp access the same bank and different addresses, it is called bank conflict.
Since bank conflict degrades read/write performance, there are known workarounds, such as shifting the boundaries of shared memory.
We show the specifications of NVIDIA Tesla V100 and A100 in Table \ref{tab:gpu-spec}.
The shared memory bandwidth is calculated assuming it is accessed without bank conflict in all SMs in one clock.
The shared memory has $12 \sim 15$ times faster bandwidth than device memory.

\subsubsection{The advantage of fewer shared memory usage}
The shared memory size that one thread block uses is one of the determining factors of occupancy, which is the max thread block size that one SM executes simultaneously.
Fewer shared memory usage means higher occupancy, which effectively hides instruction latency.
Furthermore, reducing shared memory usage can improve the L1 cache hit rate since shared memory and L1 cache resides in the same part of the chip.

\subsection{Blocking for matrix-matrix multiplication}
The number of operations of matrix-matrix multiplication $\mathbf{C} \leftarrow \mathbf{A} \cdot \mathbf{B}$ for $\mathbf{A} \in \mathbb{R}^{m \times k}, \mathbf{B} \in \mathbb{R}^{k \times n}$ is $2mnk$.
On the other hand, the sum of the number of elements in $\mathbf{A}$ and $\mathbf{B}$ is $(m + n) \times k$.
It follows $2mnk > (m + n) \times k$ in general ($m \geq 2, n \geq 2, k \geq 1$), which means that the number of operations is larger than the number of data.
Thus, data can be reused during the computation.
When computing the matrix-matrix multiplication on device  memory, we copy the sub-matrices of each input matrix from device memory to shared memory.
Then compute the matrix-matrix multiplication of these sub-matrices on shared memory to reduce the device memory footprint using data reusability.
This method of reducing the low-bandwidth memory footprint by utilizing the memory hierarchy is called ``blocking''.
The registers are also used for blocking the shared memory.
In this paper, we denote the blocking size $(m_b, n_b, k_b)$ as the size of blocking size for sub-matrices matrix-matrix multiplication $\mathbf{A}_b \cdot \mathbf{B}_b$ where the sizes of matrix $\mathbf{A}_b$ and $\mathbf{B}_b$ are $m_b \times k_b$ and $k_b \times n_b$ respectively.

\subsection{Tensor Cores}
\begin{figure}[t]
    \centering
    \includegraphics[width=\linewidth]{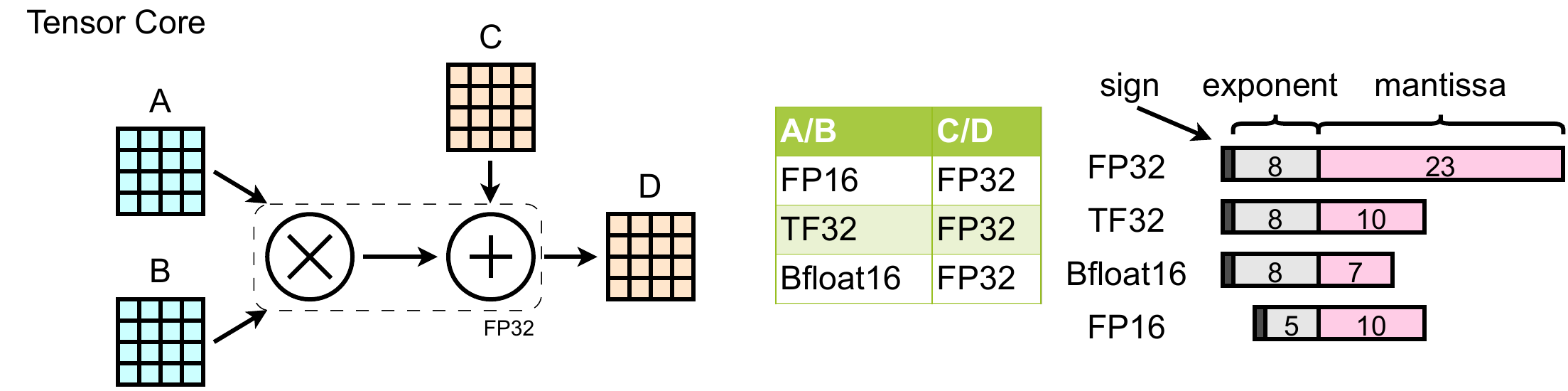}
    \caption{The input and output types of Tensor Cores on NVIDIA A100.}
    \label{fig:tc-explanation}
\end{figure}

Tensor Cores are specialized computing units for mixed-precision matrix-matrix multiplication and addition, with higher computing performance than FP16 and FP32 computing units shown in Table \ref{tab:gpu-spec}.
We show the supported input and output data types of Tensor Core in Figure \ref{fig:tc-explanation}.
We can use the TF32 (Tensor Float) data type, 8 bits of exponent and 10 bits of mantissa, and Bfloat16, 8 bits of exponent and 7 bits of mantissa, as inputs to Tensor Cores in Ampere architectures.
While TF32 has 19 bits in total, it occupies a 32-bit register and memory.
Thus, it can not be used for data compression.
\subsubsection{Programming interface}
\label{sec:programming-interface}

To use Tensor Cores in custom functions, NVIDIA provides WMMA API for C++ and Parallel Thread Execution (PTX).
When computing matrix-matrix multiplication and addition $\mathbf{D} \leftarrow \mathbf{A}\cdot\mathbf{B}+\mathbf{C}$ on Tensor Cores using WMMA API for C++, first, we copy the input matrices $\mathbf{A},\mathbf{B}$ and, $\mathbf{C}$ from memory to an array of registers called ``fragment''.
Then, we compute Matrix-Multiplication-and-Add (MMA) on the Tensor Cores and obtain the resulting $\mathbf{D}$ fragment.
The 32 threads in a warp cooperate to perform MMA operations on Tensor Cores.
Finally, we store the $\mathbf{D}$ fragment in memory.
The WMMA API provides the fragment and functions for these operations.
The fragment is a C language {\tt structure} that has an array of registers {\tt x[num\_elements]} as a member.
We show the pseudocode of simple matrix-matrix multiplication using WMMA API in Code \ref{lst:standard-wmma}.

\begin{lstlisting}[style=CStyle,caption={A simple matrix-matrix multiplication on Tensor Cores using WMMA API.},label={lst:standard-wmma}]
__device__
void matmul(float* mem_c, half* mem_a, half* mem_b) {
  using namespace nvcuda::wmma;
  fragment<matrix_a, 16, 16, 16, half, col_major> frag_a;
  fragment<matrix_b, 16, 16, 16, half, col_major> frag_b;
  fragment<accumulator, 16, 16, 16, float> frag_c;
  // Initialize an accumulator fragment
  fill_fragment(frag_c, 0.f);
  // Load matrices to fragments
  load_matrix_sync(frag_a, mem_a, ...);
  load_matrix_sync(frag_b, mem_b, ...);
  // Compute matrix-matrix multiplication
  // and accumulation on Tensor Cores
  mma_sync(frag_c, frag_a, frag_b, frag_c);
  // Store result to memory
  store_matrix_sync(mem_c, frag_c, ...);
}
\end{lstlisting}
Although the {\tt load\_matrix\_sync} function in WMMA API can generate a fragment from the device and shared memory, we consider that the shared memory is used in most cases for the following reasons:
\begin{itemize}
    \item The shared memory is used for memory blocking in matrix-matrix multiplication.
    \item The {\tt load\_matrix\_sync} function has a 128-bit alignment restriction and leading dimension size restriction.
    It is difficult to satisfy the restriction on device memory.
    %\item NVIDIA provides LDSM instruction, specialized instruction for loading a fragment from shared memory.
\end{itemize}

The fragment is regarded as a register blocking.
WMMA API specifies the blocking size of one fragment.
For instance, in the case of FP16-Tensor Core, the blocking size $(m_b,n_b,k_b)$ is one of the $(16, 16, 16),(32,8,16)$ or $(8,32,16)$.
We can use the array of fragments to increase the blocking size.

\subsubsection{Mapping between memory and fragment}
\begin{figure}[t]
    \centering
    \includegraphics[width=0.8\linewidth, bb=0 0 685.91998 522]{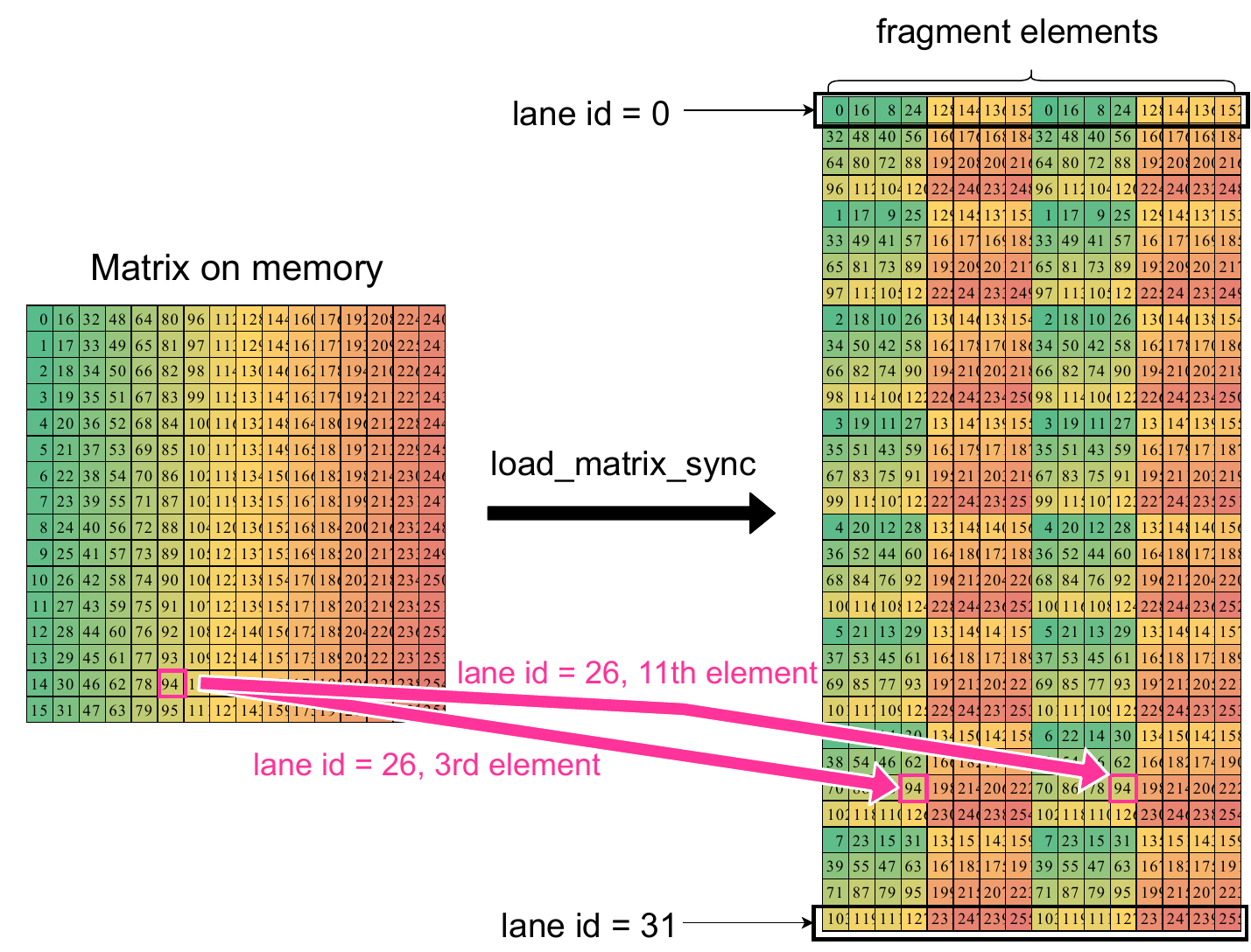}
    \caption{An example of memory-fragment mapping. The {\tt lane id} is a thread number in a warp which is calculated by {\tt (threadIdx.x \& 0x1f)}.}
    \label{fig:fragment-mapping}
\end{figure}

Each matrix element in memory is stored as an element of a fragment of some thread.
Although the mapping between memory and fragment elements is not public, we can investigate it \cite{jia_dissecting_2018,li_tcfft_2021,dakkak_accelerating_2019}.
This mapping depends on the type, memory layout, etc, of the matrix.
We use Code \ref{lst:investigate-mapping} to investigate the mapping and show an example of the mapping in Figure \ref{fig:fragment-mapping}.
\begin{lstlisting}[style=CStyle,caption={A kernel function to investigate the memory-fragment mappings.},label={lst:investigate-mapping}]
template <class Use, class Layout, class T>
__global__ void investigate_mapping() {
  __shared__ T smem[];
  // initialize smem
  for (i = 0; i < 16 * 16; i++) smem[i] = i;
  fragment<Use, 16, 16, 16, T, Layout> frag;
  load_matrix_sync(frag, smem, ...); // WMMA API
  for (i = 0; i < 32; i++) {
    if (threadIdx.x == i){
        for (j = 0; j < frag.num_elements; j++) {
            // Print the mapping
            printf("%d,", (int)frag.x[j]);}
        printf("\n");}
    __syncwarp();}}
\end{lstlisting}

\subsubsection{WMMA API for PTX}
The WMMA API for PTX provides two types of instructions: 1) wmma instructions and 2) mma instruction.
The WMMA API for C++ functions calls wmma instructions using inline assembly.
The wmma instructions include functionality for loading and storing fragments and MMA operation.
On the other hand, mma instruction only includes MMA operation.
Thus, when using mma instruction, we must manually load fragments from memory.
The mapping is available on CUDA developer documentation.
There is a difference between the wmma instructions and the mma instruction regarding register usage.
When using wmma instructions, one element in a matrix is kept by two elements in a fragment in 32 threads in a warp.
On the other hand, when using mma instruction, one element in a matrix is kept by only one element in a fragment in 32 threads in a warp without duplication.
Thus, the mma instruction computes MMA operation using fewer registers than the wmma instructions.

\section{The balance of Tensor Cores performance and shared memory bandwidth}
Although the shared memory bandwidth is higher than device memory, the computing performance of the Tensor Cores is high, and its Bytes-per-Flops (B/F) ratio is calculated to be $0.06 \sim 0.12$ from Table \ref{tab:gpu-spec}.
This value is similar to the ratio between the FP32 computing unit and device memory ({0.06 $\sim$ 0.10}).
In the case of the FP32 computing unit and device memory, the memory blocking using shared memory reduces global memory access and alleviates the problem of this small B/F ratio.
Similarly, in the case of shared memory and Tensor Cores, it is important to reduce shared memory accesses to take advantage of high computational performance.

Now, we analyze a matrix-matrix multiplication on Tensor Cores using the roofline model.
The input matrices $\mathbf{A}$ and $\mathbf{B}$ are FP16, $\mathbf{C}$ and $\mathbf{D}$ are FP32 stored in the shared memory.
We load the sub-matrices of each input matrix as fragments $\mathbf{A}_\text{reg}$ and $\mathbf{B}_\text{reg}$ for register blocking.
The register blocking size is $(n, n, n)$.
We show the roofline model of computing $\mathbf{D}_\text{reg} \leftarrow \mathbf{A}_\text{reg} \cdot \mathbf{B}_\text{reg} + \mathbf{C}_\text{reg}$ in Figure 1.
The Arithmetic Intensity (AI) is calculated as follows:
\begin{equation}
    \label{eq:simple-mma-ai}
    \text{AI} = \frac{2n^3}{(n^2 + n^2) \text{sizeof(FP16)} + (n^2 + n^2) \text{sizeof(FP32)}} = \frac{n}{5}.
\end{equation}
As the size of register blocking size increases, we can utilize the performance of Tensor Cores more.
However, the number of registers is finite, and the registers spill to local memory when using more than 256 registers per thread.
The number of 32-bit registers required for the blocking is calculated as follows assuming the mma instruction is used and each element in a matrix is stored by only one element of a fragment without duplication.
\begin{equation}
    \label{eq:simple-mma-reg}
    \text{nRegs} = ((\underbrace{n^2}_{\mathbf{A}_\text{reg}} + \underbrace{n^2}_{\mathbf{B}_\text{reg}}) \times \frac{1}{2}
    + \underbrace{n^2}_{\mathbf{C}_\text{reg}}) / \text{warpSize}
    = \frac{1}{16} n^2.
\end{equation}
For instance, in the case of $n=64$, the number of required registers is 256, and the registers spill to local memory.
Therefore, we need to reduce the shared memory access not by increasing the register blocking size.
Furthermore, the Tensor Cores performance has been improved more than the shared memory bandwidth on NVIDIA A100 compared to V100.
This can be seen from the fact that the AI value at the boundary between the memory bandwidth and the computational performance bound is smaller for A100 than for V100.

\begin{figure}[t]
    \centering
    \includegraphics[width=0.9\linewidth]{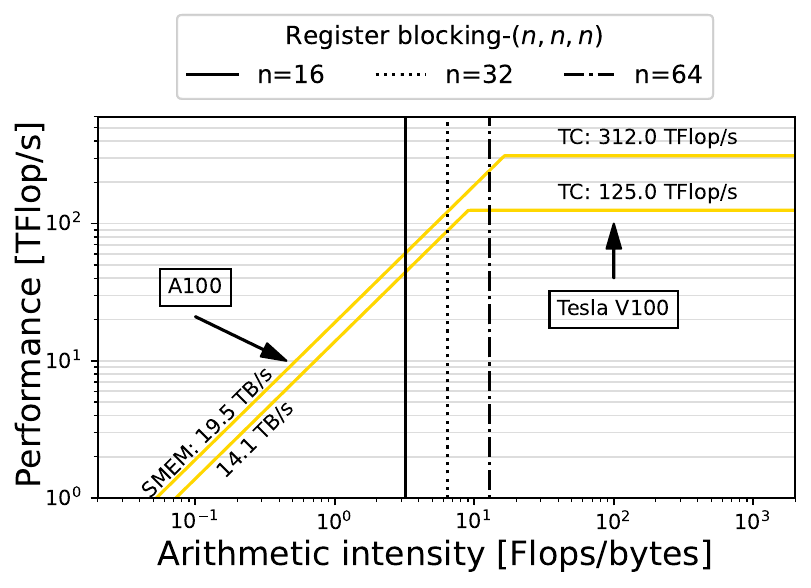}
    \caption{The arithmetic intensity of matrix-matrix multiplication for each size of register blocking blocking-$(n, n, n)$.%
    }
    \label{fig:full-nocor-roofline}
\end{figure}

\section{WMMA API extension library}
To leverage the high Tensor Cores performance, it is necessary to supply matrix data to Tensor Core with sufficient throughput.
However, due to the limited functionality of the WMMA API, the throughput improvements that can be made using only the WMMA API are limited.
Therefore, we implement a WMMA API Extension library (WMMAe) to reinforce the functionality of WMMA API.
The WMMAe consists of the following two components:
\begin{enumerate}
    \item Primitive functions
    \item SGEMM emulation on Tensor Cores using Error Correction method (WMMAe-TCEC)
\end{enumerate}
In this section, we show the functionality of these components and evaluate the performance improvement compared to only using WMMA API.
We use NVIDIA A100 40GB SXM4 and NVIDIA V100 16GB PCIe GPUs for the evaluations.

\subsection{Primitive functions}
We can generate a fragment of a matrix in which all elements are the same value without shared memory access using {\tt fill\_fragment} function in WMMA API.
On the other hand, to generate fragments of other matrices, it is necessary to explicitly store the matrix in shared memory and load it using {\tt load\_matrix\_sync} function in WMMA API.
Now, we consider the matrices that have some structural rules.
For instance, when performing scan operations using matrix-vector multiplication, we need an upper triangular matrix $\mathbf{U}$ in which all non-zero elements are one.
Then, we perform a scan operation to an array $\begin{bmatrix}a_{0} & a_{1} & \cdots & a_{n - 1}\end{bmatrix}$ using $n \times n$ matrix $\mathbf{U}$ as follows:

\begin{eqnarray}
    \begin{bmatrix}a_{0} \\ a_{1} \\ \vdots \\ a_{n-1}\end{bmatrix}^\top \cdot \mathbf{U} =
    \label{eq:scan-mat}
    \begin{bmatrix}a_{0} \\ a_{1} \\ \vdots \\ a_{n-1}\end{bmatrix}^\top \cdot
    \begin{bmatrix}
        1 & 1 & 1 & \cdots & 1 \\
        0 & 1 & 1 & \cdots & 1 \\
        0 & 0 & 1 & \cdots & 1 \\
        \vdots & \vdots & \vdots & \ddots & \vdots \\
        0 & 0 & 0 & \cdots & 1
    \end{bmatrix}
    =\begin{bmatrix}a_{0} \\ \sum_{i=0}^{1} a_{i} \\ \vdots \\ \sum_{i=0}^{n-1} a_{i} \end{bmatrix}^\top. \nonumber
\end{eqnarray}
The structural rule for the $(i, j)$ element of the matrix $\mathbf{U}$ is as follows:
\begin{equation}
\label{eq:u-rule}
    u_{i,j} = \left\{
\begin{array}{ll}
    1 & i \leq j \\
    0 & \text{Otherwise}
\end{array}
\right.
\end{equation}
Dokkak \textit{et al.} utilize the rule for generating the fragment of the matrix without storing it explicitly in shared memory.
We generalize the functionality and provide functions for generating a fragment of any matrix from its structural rule: {\tt foreach\_ij} and {\tt map}.

\subsection{Primitive function : {\tt foreach\_ij}}
The {\tt foreach\_ij} function calculates the mapping between matrix element position $(i, j)$ and fragment indices and gives them to a given lambda function.
In the lambda function, we calculate the value of the $(i, j)$ element of the matrix and set it to the fragment using the given mapping information.
For instance, we show a pseudocode for generating the matrix U fragment by the rule in Eq (\ref{eq:u-rule}) in Code \ref{lst:foreach-ij}.
Strictly speaking, since one element in a matrix is kept by two fragment elements when using WMMA API for C++, {\tt foreach\_ij} function gives the list of fragment element indices to the lambda function.
However, in this pseudocode, we simplify the argument of the lambda function as only one fragment index is given.
By using this function, we can generate a fragment of any matrix from its structural rule without storing it in shared memory.
\begin{lstlisting}[style=CStyle,caption={Generating the matrix $\mathbf{U}$ fragment from the structual rule in Eq. (\ref{eq:u-rule}) using WMMAe {\tt foreach\_ij} function.},label={lst:foreach-ij}]
fragment<16, 16, 16> frag;
foreach_ij<decltype(frag)>(
  // The lambda function to set each fragment elements
  [&](fid, i, j) {
      if (i <= j) frag.x[fid] = 1;
      else frag.x[fid] = 0;
    });
\end{lstlisting}

\subsubsection{Performance evaluation}
\begin{figure}[t]
    \centering
    \includegraphics[width=\linewidth]{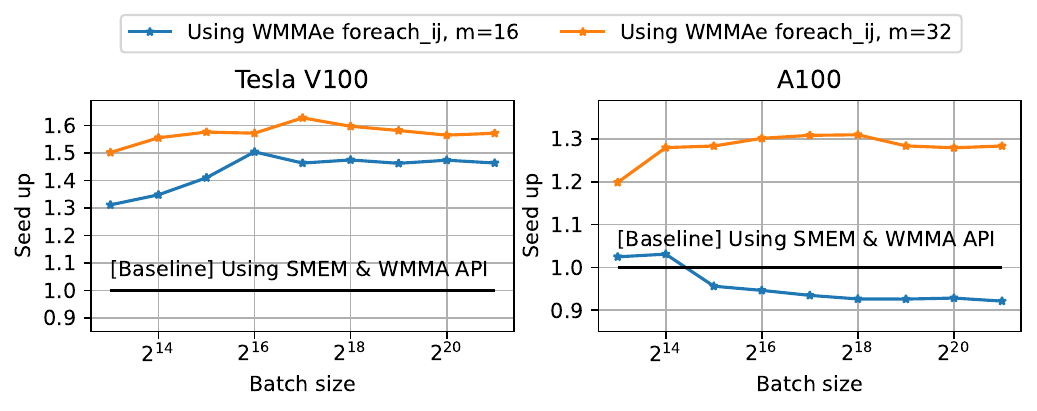}
    \caption{The performance evaluation of {\tt foreach\_ij} function using batched Householder benchmark, where we multiply a $m \times m$ Householder matrix $\mathbf{H}$ with an input matrix $\mathbf{A}$ using Tensor Cores.}
    \label{fig:householder-performance}
\end{figure}
We use a batched Householder transformation benchmark for evaluating the performance improvement by {\tt foreach\_ij} function.
The Householder transformation is one of the orthogonal transformations used for QR factorization etc.
This transformation is calculated as follows for a $n \times n$ Householder matrix $\mathbf{H}$, $m \times k$ input matrix $\mathbf{A}$:
\begin{equation}
    \label{eq:householder}
    \mathbf{H}\cdot \mathbf{A} = \left(\mathbf{I}_m - 2 \mathbf{v}^\top\mathbf{v}\right) \cdot \mathbf{A},
\end{equation}
where $\mathbf{v}$ is a $m$-dimensional identity vector and $\mathbf{I}_m$ is a $m \times m$ identity matrix.
In this benchmark, we explicitly compute the Householder matrix $\mathbf{H}$ from $\mathbf{v}$ and multiply it by $\mathbf{A}$.
This computation is performed for $b$ (batch size) FP16 input matrices $\mathbf{A}_i$ and FP16 vectors $\mathbf{v}_i$.
To obtain the baseline performance, we implemented the batched Householder transformation, which stores the Householder matrix in shared memory and loads it using the WMMA API function.
Then the multiplication of $\mathbf{A}$ and $\mathbf{H}$ is performed on Tensor Cores.
We show a speed-up ratio using WMMAe in Figure \ref{fig:householder-performance}.
We can see that the performance is improved using {\tt foreach\_ij} on V100 GPU in both cases.
%{\tt foreach\_ij}を用いた実装では、$m=16$の場合はコード1のように実装した。
On the other hand, for $m=16$ on A100, the implementation using {\tt foreach\_ij} has a lower performance compared to the baseline.
In this case, the pseudocode of the implementation is shown in Code \ref{lst:householder-foreach-16}.
\begin{lstlisting}[style=CStyle,caption={Generating a $16 \times 16$ Householder matrix fragment using {\tt foreach\_ij}}.,label={lst:householder-foreach-16}]
fragment<16, 16, 16> frag;
foreach_ij<decltype(frag)>(
  [&](fid, i, j) {
    auto elm = v[i]*v[j]*(-2);
    if (i==j) elm += 1
    frag.x[fid] = elm;
  });
\end{lstlisting}
In this code, the cost of the mapping calculation is higher than the cost of storing the matrix explicitly in shared memory, which might be the reason for the low performance.
Whereas, for $m=32$ on A100, the implementation using {\tt foreach\_ij} has higher performance than the baseline.
In this case, the pseudocode of the implementation is shown in Code \ref{lst:householder-foreach-32}.
\begin{lstlisting}[style=CStyle,caption={Generating an array of fragments for $32 \times 32$ Householder matrix using {\tt foreach\_ij}.},label={lst:householder-foreach-32}]
fragment<16, 16, 16> frag[2 * 2];// 32x32 matrix
foreach_ij<decltype(*frag)>(
  [&](fid, i, j) {
    for (unsigned bi = 0; bi < 2; bi++) {
      for (unsigned bi = 0; bi < 2; bi++) {
        auto elm = v[i+bi*16] * v[j+bj*16]*(-2);
        if (i==j) elm += 1
        frag[bi+bj*2].x[fid] = elm;
      }}});
\end{lstlisting}
For the $32 \times 32$ matrix fragment, we used a $2 \times 2$ array of fragments holding matrices of size $16 \times 16$.
The elements of all the fragments are set in a single {\tt foreach\_ij} function.
This means that four fragments are generated in one mapping calculation, and the cost of the mapping calculation is relatively lower than that of the $m=16$ case.
Thus, we consider that reusing the mapping calculation among several fragments is important to speed up the use of the {\tt foreach\_ij} function.

\subsection{Primitive function : {\tt map}}
The map function takes the position $(i, j)$ of an element of the matrix as an argument and returns a pair {\tt (lid, fid)} of the thread number (lane id; {\tt lid}) in a warp and the element number of the fragment holding this element.
Using this function, we can manipulate any $(i, j)$ element of the matrix as a fragment.
For instance, Code \ref{lst:map-example} sets the $(i, j)$ element of a matrix $\mathbf{A}$, which is held as a fragment, to 1.

\begin{lstlisting}[style=CStyle,caption={Setting $(i, j)$-element of a matrix held as fragment using WMMAe {map} function.},label={lst:map-example}]
fragment frag_a;
unsigned lid, fid;
// Calculate lid and fid from matrix position (i, j)
map<decltype(frag)>(lid, fid /*=2*/, i, j);
// Set 1
if ((threadIdx.x & 0x1f) == lid) {
  frag_a.x[fid] = 1;
}
\end{lstlisting}

\subsubsection{Performance evaluation}
\begin{figure}[t]
    \centering
    \includegraphics[width=\linewidth]{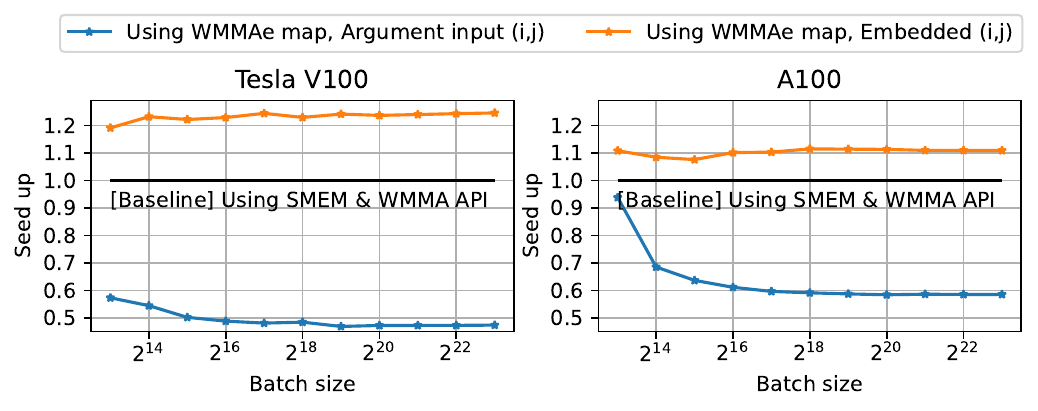}
    \caption{
    The performance evaluation of {\tt map} function using batched Given's rotation benchmark.
    The ``Argument input (i,j)" means that the parameter (i, j) for Given's rotation matrix is set through kernel function arguments, and ``Embedded (i,j)" means that these parameters are set in compile-time.}
    \label{fig:givens-performance}
\end{figure}
We define a batched  Given's rotation benchmark to evaluate the performance improvement by the map function.
The Given's rotation is a rotation operation for a vector and matrix and is used for QR factorization etc.
The definition of Given's rotation for a matrix $\mathbf{A}$ is as follows:
\begin{equation}
    \mathbf{G}(i, j, \theta)\cdot\mathbf{A},
\end{equation}
where
\begin{equation*}
    %\label{eq:givens-rotation}
    \begin{blockarray}{cccccccccc}
    \begin{block}{c[ccccccc]cc}
    &1 &        &        &        &        &        &   &              & \\
    &  & \ddots &        &        &        &        &   &              & \\
    &  &        & c      & \cdots & -s     & \cdots &   & i \text{-th} &\\
    \mathbf{G}(i, j, \theta) = &  &        & \vdots & \ddots & \vdots &        &   &              &\\
    &  &        & s      & \cdots & c      & \cdots &   & j \text{-th} &\\
    &  &        & \vdots &        & \vdots & \ddots &   &              &\\
    &  &        &        &        &        &        & 1 &              &\\
    \end{block}
    &  & & i\text{-th} & & j\text{-th}& & &
    \end{blockarray},
\end{equation*}
$c = \cos{\theta}$ and $s = \sin{\theta}$.
In this benchmark, Given's rotation operations for $b$ FP16 input matrices $\mathbf{A}_k$ are performed by multiplying by $\mathbf{G}(i, j, \theta_k)$ in parallel.
The $i$ and $j$ are fixed in all calculations, and there are two ways to fix them as follows:
1) Specify them as arguments of the kernel function. 2) Embed them in the kernel function.
When generating a fragment of matrix $\mathbf{G}$ using the map function, first, all elements in the fragment are filled with zeros by WMMA API {\tt fill\_fragment} function.
Then set $1, s$, and $c$ at each position using the {\tt map} function.
To obtain the baseline performance, the matrix $\mathbf{G}$ is explicitly stored in shared memory and loaded using a WMMA API function.
We show a speed-up ratio by the mapping function in Figure \ref{fig:givens-performance}.
When $i$ and $j$ are given as arguments of the kernel function, it is slower than the baseline implementation.
On the other hand, when $i$ and $j$ are embedded in the kernel function, then the baseline implementation.
When $i$ and $j$ are embedded in the kernel function, compiler optimization reduces the computing amount of mapping calculation and required registers at runtime.

\subsection{WMMAe-TCEC}
When computing single-precision matrix-matrix multiplication on Tensor Cores, we need to convert input matrices to FP16 ones.
This conversion results in a loss of accuracy in the resulting matrix.
Markidis \textit{et al.} \cite{markidis_nvidia_2018} proposed a method for single-precision matrix multiplication using Tensor Cores with error correction.
However, the accuracy of their method does not match the single-precision.
In our previous research, we improve the accuracy and reduce the computation complexity of their method \cite{ootomo_recovering_2022}.
In our method, they compute the single-precision matrix-matrix multiplication $\mathbf{C}_\text{F32}=\mathbf{A}_\text{F32} \mathbf{B}_\text{F32}$ as follows:
\begin{eqnarray}
\label{eq:corr-1-1024}
\mathbf{A}_\text{F16} &\leftarrow& \text{toFP16}\left( \mathbf{A}_\text{F32} \right) \\
\Delta\mathbf{A}_\text{F16} &\leftarrow& \text{toFP16}\left( \left(\mathbf{A}_\text{F32} - \text{toFP32}\left(\mathbf{A}_\text{F16}\right)\right) \times 2^{11}\right) \nonumber
\label{eq:corr-2-1024}
\\
\label{eq:corr-3-1024}
\mathbf{B}_\text{F16} &\leftarrow& \text{toFP16}\left( \mathbf{B}_\text{F32} \right) \\
\Delta\mathbf{B}_\text{F16} &\leftarrow& \text{toFP16}\left( \left( \mathbf{B}_\text{F32} - \text{toFP32}\left(\mathbf{B}_\text{F16}\right)\right) \times 2^{11} \right) \nonumber 
\label{eq:corr-4-1024}
\\
\label{eq:corr-5-1024-reduce}
\mathbf{C}_\text{F32} &\leftarrow& \mathbf{A}_\text{F16} \mathbf{B}_\text{F16}
+ \left(\Delta\mathbf{A}_\text{F16} \mathbf{B}_\text{F16}  + \mathbf{A}_\text{F16} \Delta\mathbf{B}_\text{F16}\right) /2^{11}\nonumber,
\end{eqnarray}
where toFP16 and toFP32 are the conversion to FP16 and FP32, respectively.
We improve the matrix-matrix multiplication accuracy by avoiding the rounding inside Tensor Cores, RZ, and achieve the same accuracy with FP32 SIMT Core computation.
Although we have included our method in NVIDIA CUTLASS and evaluated the accuracy, performance, and power consumption in the previous paper, the matrix-matrix multiplication is inside various linear algebra algorithms, and we would like to use the computation inside custom kernel functions.
Therefore, we provide functionality for using this method inside a custom kernel function.

\begin{figure}[t]
    \centering
    \includegraphics[width=\linewidth]{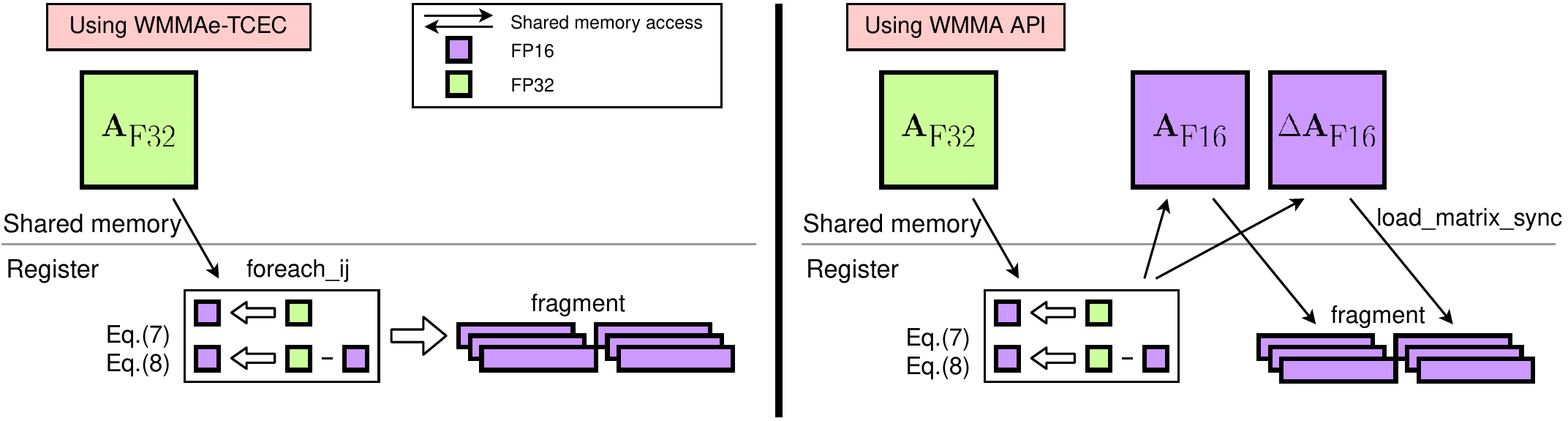}
    \caption{The comparison of data flow between using WMMA API and WMMAe.
    Here we load fragments for SGEMM emulation on Tensor Cores using error correction without additional shared memories $\mathbf{A}_\text{F16}$ and $\Delta\mathbf{A}_\text{F16}$, that are required when using WMMA API.}
    \label{fig:mma_f32-load}
\end{figure}
To compute the Eqs. (\ref{eq:corr-1-1024})-(\ref{eq:corr-5-1024-reduce}) using WMMA API for C++, we need to store the matrices $\mathbf{A}_\text{F16}, \Delta \mathbf{A}_\text{F16}$ in the shared memory explicitly since the mapping function {\tt load\_matrix\_sync} in WMMA API only makes the {\tt fragment} from memory as shown in the top of Figure \ref{fig:mma_f32-load}.
On the other hand, we can avoid the explicit storing by {\tt foreach\_ij} function in WMMAe.
Using this function, we implement WMMAe-TCEC, which reduces the memory footprint and provides the error correction computation with the same interface as WMMA API.
The WMMAe-TCEC includes a function for generating the fragments of $\mathbf{A}_\text{F16}$ and $\Delta \mathbf{A}_\text{F16}$ directly from the input matrix $\mathbf{A}_\text{F32}$ shown in the bottom of Figure \ref{fig:mma_f32-load}.
We can use WMMAe-TCEC just by changing the matrix data types and the namespace in Code 1 from
    {\tt nvcuda::wmma}
to
    {\tt mtk::wmma::tcec}.

Moreover, since the WMMAe-TCEC adopts a policy-based design, we can change the following backward computation by only changing the policy, which is specified as an optional template parameter of the fragment.
\begin{itemize}
    \item
    Tensor Core instruction: Use the wmma instructions or mma instruction.
    \item
    Error correction: Enable or disable.
    \item
    Use Tensor Core or software systolic array \cite{chen_versatile_2019}.
\end{itemize}
Using this feature, we can evaluate the effect of the error correction method easily.

\subsubsection{Theoretical performance analysis}
\begin{figure}[t]
    \centering
    \includegraphics[width=0.9\linewidth]{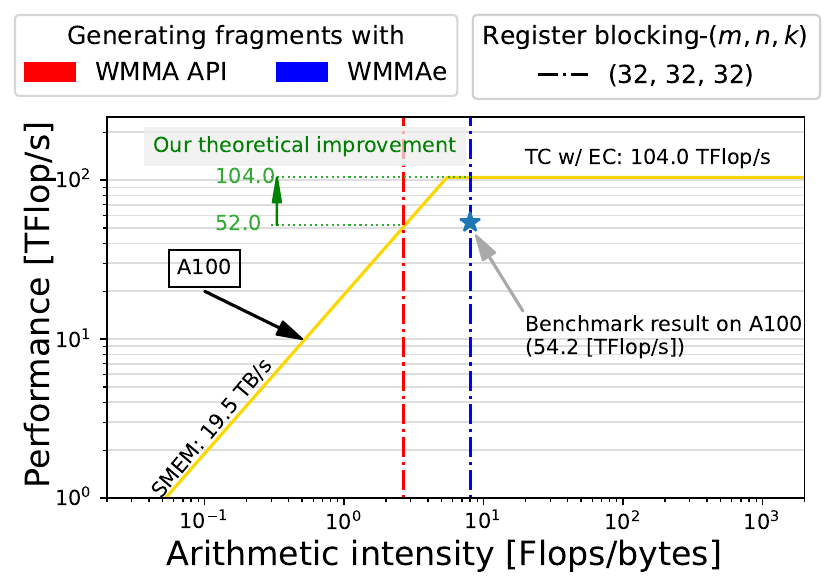}
    \caption{The arithmetic intensity of SGEMM emulation on Tensor Cores using error correction method. %
    The peak performance is calculated by dividing the theoretical peak performance of FP16-TC in Table \ref{tab:gpu-spec} by 3 since we need 3 times matrix-matrix multiplication in Eq. (\ref{eq:corr-5-1024-reduce}).}
    \label{fig:full-roofline}
\end{figure}
We show the AI of matrix-matrix multiplication with error correction that we used for the performance evaluation in Figure \ref{fig:full-roofline}.
By using WMMAe-TCEC, we can increase the AI and improve the theoretical computing performance bounded by the shared memory bandwidth.
Although we can increase the AI by increasing the size of register blocking, the number of registers that one thread can use is limited by the hardware.
%In the error correction computation, we need $\times 2$ registers compared to without error correction since we need to keep $\Delta \mathbf{A}_\text{F16}$, and $\Delta \mathbf{B}_\text{F16}$ fragments in Eq.(\ref{eq:corr-2-1024}) and Eq.(\ref{eq:corr-4-1024}).
For instance, in the case of $(m, n, k) = (32, 32, 32)$, which is used in our benchmark evaluation, we need 128 32-bit registers to keep the fragments, which amounts to 50\% of registers that one thread can use.
The registers are used not only for fragments but also for memory access offset calculations and other floating-point value operations such as eq. (\ref{eq:corr-2-1024}).
Reducing the number of required registers can improve the throughput since it can improve occupancy.
And when the number of required registers exceeds the hardware limitation, the device memory is used instead, which results in performance degradation.
Therefore, increasing the AI without increasing the register blocking size is advantageous.

\subsubsection{Performance evaluation}
\begin{figure}[t]
    \centering
    \includegraphics[width=\linewidth]{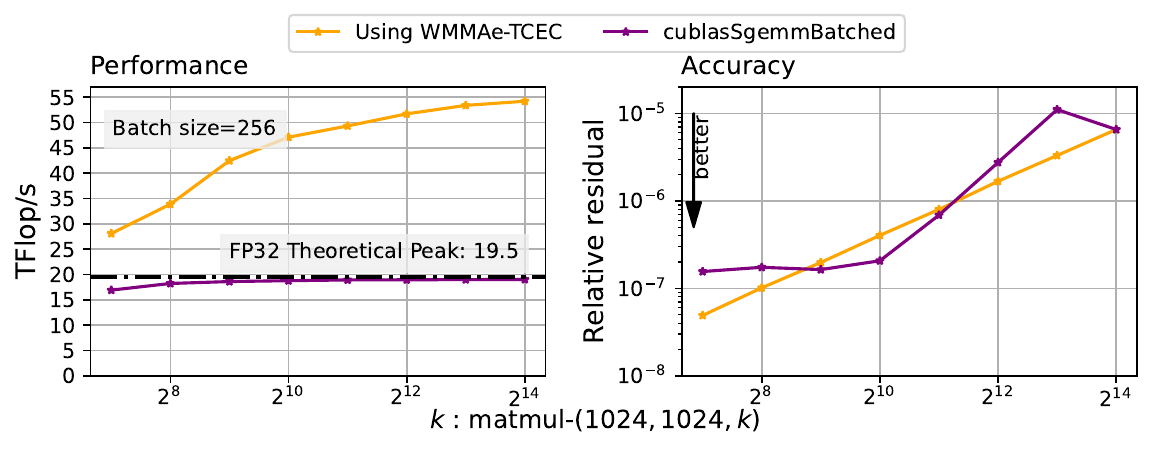}
    \caption{The throughput and accuracy evaluation of batched SGEMM using WMMAe-TCEC.}
    \label{fig:bgemm-evaluation}
\end{figure}
We use a batched matrix-matrix multiplication benchmark to evaluate the performance and accuracy of the WMMAe-TCEC.
In this benchmark, we compute $256$ matrix-matrix multiplications $\mathbf{A}_i \cdot \mathbf{B}_i$ where each $\mathbf{A}_i$ and $\mathbf{B}_i$ are $1024 \times k$ and $k \times 1024$ FP32 matrices.
Then, we calculate the computing performance from the computing time $t$ [s] as $(2\times 1024 \times 1024 \times k / t)$ [Flop/s], and a max relative error for the accuracy.
We show the performance and accuracy comparison between our implementation using WMMAe-TCEC and cuBLAS batched SGEMM function in Figure \ref{fig:bgemm-evaluation}.
In our implementation, we use the mma instruction, and the shared memory and register blocking sizes are $(128, 128, 32)$ and $(32, 32, 32)$, respectively.
We found this blocking size using a grid search that experimentally maximizes the throughput on NVIDIA A100 (40GB, SXM4) GPU.
The outcome of our evaluation shows that our implementation achieves 54.2 [TFlop/s], which outperforms the theoretical peak performance of FP32 on NVIDIA A100, while the accuracy remains the same with cuBLAS SGEMM.
The achieved throughput is larger than the throughput of SGEMM emulation that we have achieved using the NVIDIA CUTLASS library (51 TFlop/s) in our previous paper \cite{ootomo_recovering_2022}.
According to the roofline model, when we only use WMMA API, the theoretical peak performance for our chosen register blocking size is limited to 52.0 TFlop/s bounded by the shared memory bandwidth.
Therefore, the achieved throughput can not be achieved without reducing the shared memory footprint that our library does.
However, by using WMMAe, we improved the theoretical peak performance of this method to 104.0 TFlop/s by reducing the shared memory footprint.
Since the achieved efficiency is only 52\% of the theoretical peak performance, we believe there is room for improving the throughput.

We summarize the advantages of WMMAe-TCEC as follows:
\begin{itemize}
    \item It provides an interface for the single-precision emulation method on Tensor Cores, which has the same interface as NVIDIA WMMA API.
    \item It improves the theoretical peak performance of matrix-matrix multiplication with error correction by reducing shared memory footprint without increasing register usage.
    \item It reduces the shared memory usage required to store the fragments of FP16 matrices when using only WMMA API.
    \item It is proved to outperform the FP32 theoretical peak performance on NVIDIA A100 experimentally while the accuracy remains the same with FP32 computation.
\end{itemize}

\section{Conclusion}
We have investigated a simple matrix-matrix multiplication on Tensor Cores by roofline model and found that reducing the shared memory footprint is necessary to fully exploit the high throughput of Tensor Cores.
To reduce the footprint, we implement a WMMA API extension library which allows us to generate fragments flexibly.
This library is open-source and available on GitHub.
We show that this library can improve the computing throughput on Tensor Cores.
Furthermore, we improve the theoretical peak performance of single precision matrix-matrix multiplication emulation on Tensor Cores, which is bounded by the shared memory bandwidth when using only WMMA API.
Then, we provide this functionality with the same interface as WMMA API.
We also show that this functionality can outperform the FP32 theoretical peak performance on NVIDIA A100 GPU.
We believe such a faster data supply is necessary to maximize the use of high-speed matrix multiplication units in future architectures.

\begin{acks}
This work was partially supported by JSPS KAKENHI JP22H03598 and JP21J14694.
This work was partially supported by "Joint Usage/Research Center for Interdisciplinary Large-scale Information Infrastructures" in Japan (Project ID: jh220022-NAHI)
\end{acks}

\bibliographystyle{ACM-Reference-Format}
\bibliography{references}

\end{document}